# Phase Coherent Transport in Two-Dimensional Tellurium Flakes

*Mohammad Hafijur Rahaman[1], Nathan Sawyers[1], Mourad Benamara[2], Trudie Culverhouse[3,4], Repaka Maheswar[3], Qiyuan He[5], Hugh Churchill[1,6], Dharmraj Kotekar Patil[1,6]*

[1]Department of Physics, University of Arkansas, Fayetteville, AR 72701 USA

[2]Institute for Nano Science and Engineering, University of Arkansas, Fayetteville, AR 72701 USA

[3] Institute of Materials Research and Engineering, Agency for Science Technology and Research, (A*STAR) Singapore 138634, Republic of Singapore 138634

[4]Department of Chemistry, School of Natural Sciences, University of Manchester, Manchester M13 9PL, United Kingdom

[5]Department of Materials Science and Engineering, City University of Hong Kong, 83 Tat Chee Avenue, Kowloon, Hong Kong, China

[6]MonArk NSF Quantum Foundry, University of Arkansas, Fayetteville, AR 72701 USA

**ABSTRACT**

Elemental tellurium (Te) is a compelling van der Waals material due to its interesting chiral crystal structure and predicted topological properties. Here, we report the fabrication and comprehensive quantum transport study of devices based on Te flakes with varying thicknesses. We demonstrate a hole mobility reaching up to 1000 cm²/V·s in a 17 nm thick flake at 30 Kelvin. At deep cryogenic temperatures (< 50mK), the transport characteristics transition from Coulomb blockade in the low carrier density regime to pronounced Fabry-Pérot (F-P) interference at higher densities. Notably, the visibility of these F-P oscillations is significantly enhanced in the thinner flake device. The application of a magnetic field reveals a clear Zeeman splitting of the conductance peaks. The rich variety of quantum transport phenomena observed underscores the high quality of our thin Te flakes and establishes them as a promising platform for exploring novel physics and device concepts, such as topological superconductivity and low-power spintronic applications.

**INTRODUCTION**

The exploration of quantum phenomena in two-dimensional (2D) van der Waals materials has been a central and highly active theme in modern condensed matter physics. Seminal discoveries in materials like graphene[1] and transition metal dichalcogenides[2] have not only revealed a host of novel physical effects[3–6] but have also opened pathways for next-generation electronic and optoelectronic devices.[7–13] A key frontier in this field is the search for and characterization of materials with non-trivial electronic topology, which promise to host exotic quasiparticles and protected quantum states. This pursuit has driven research beyond established 2D systems (e.g. graphene and TMDCs) toward elemental materials whose unique crystal structures and electronic properties may offer new platforms for investigating fundamental physics and realizing novel device functionalities.

Within this landscape, elemental tellurium (Te) has recently emerged as a uniquely compelling candidate. Its chiral, one-dimensional chain-like crystal structure[14,15] (Figure 1a) gives rise to theoretically predicted Weyl-semimetal characteristics and the potential for non-trivial topological electronic states.[16,17] While the properties of bulk Te are well-documented,[18–21] realizing its full potential for quantum device applications requires the fabrication of high-quality, thin crystalline flakes where quantum phenomena are enhanced and can be systematically investigated. While mechanical exfoliation and chemical vapor deposition (CVD) are standard for producing 2D films from van der Waals materials, creating large-area thin 2D tellurium has been challenging. Previous attempts to grow tellurium nanostructures resulted in either quasi-1D forms or relatively small 2D nanostructures, unsuitable for transport studies. [22–24] A recently introduced liquid-based synthesis method, however, successfully produces large-scale 2D tellurene films.[25,26] With dimensions exceeding tens of microns and variable thicknesses, these high-quality nanofilms have been validated and present a promising new platform for investigating both electronic devices and magneto-transport properties. The recent discovery of emergent topological phenomena in elemental tellurium, such as Weyl nodes and signatures of the quantum hall effect, has sparked significant interest in leveraging its chiral structure to explore non-trivial electronic properties.[27–30] Here, we report on the successful fabrication and comprehensive low-temperature transport study of field-effect devices based on Te flakes with varying thicknesses. We demonstrate relatively high carrier mobility that is dependent on flake thickness and observe a rich spectrum of quantum transport features, including a clear observation of Coulomb blockade to pronounced Fabry-Pérot interference (F-P). F-P interference is often observed in highly crystalline materials (e.g. graphene, group III-V and IV semiconductors, etc)[31–33] reflecting high crystal quality of our thin Te flakes and its suitability for quantum transport studies.

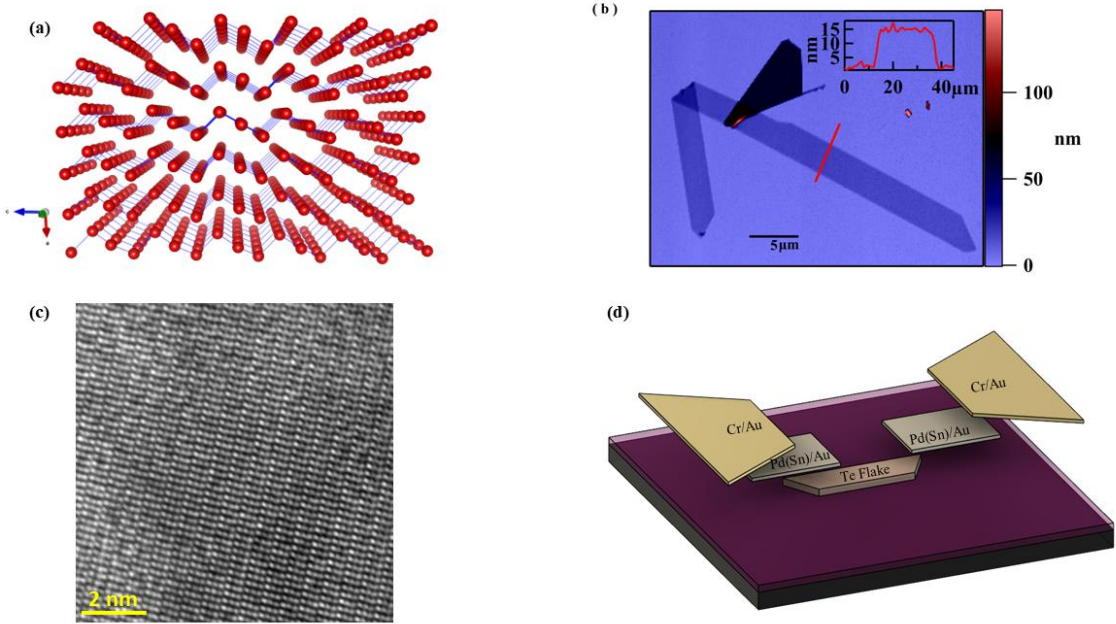

*Figure 1. (a) Schematic of the Te crystal exhibiting helical Te chains.(b) Atomic force microscope micrograph of a Te flake of thickness 13 nm. inset shows the linecut of the flake height. Scale bar in inset is 5μm. (c) Transmission electron micrograph of Te flake showing atomic structure. (d) Schematic of the device structure, including Pd (Sn) / Au contacts and a silicon.*

**Experimental Results**

**CRYSTAL CHARACTERISATION AND DEVICE FABRICATION**

Tellurene (Te) flakes were synthesized using a hydrothermal method[26] and transferred onto heavily doped p-type silicon substrates capped with a 285 nm $SiO_2$ layer, which served as a global back gate to modulate carrier density in the Te channel. The as-grown Te flakes typically display a trapezoidal geometry (Figure 1b) with the long edge aligned along the helical axis of the crystal, aiding the alignment with device orientation. Structural characterization of 2D tellurium flakes is performed using transmission electron microscopy. Figure 1c shows a high-angle annular dark-field scanning transmission electron microscopy (HAADF-STEM) image. The resolved the atomic structure shows helical tellurium chains with threefold screw symmetry along the [0001] direction. Measured interplanar spacings of 2.3 Å and 6.0 Å correspond to the ($\bar{1}210$) and (0001) lattice planes of 2D tellurium, respectively.

Two-terminal field-effect transistor devices were defined via maskless photolithography, and metal contacts, either tin (Sn) or palladium (Pd) were deposited using electron beam evaporation. A schematic of the different device layers is shown in Figure 1d. All devices in this study employed Te flakes with thicknesses ranging from ~13 nm to 60 nm. The device measurements presented in the main text of this manuscript display device measurements performed on flakes with thicknesses of 17 nm and 16 nm, while measurements on other thicknesses are presented in supplementary information (Supplementary Figure S1).

Devices fabricated using thinner flakes (13–20 nm) exhibited a strong gate tunability with on/off ratios exceeding $10^3$ (limited by the off-state current in the measurement set-up) (Figure 2a), while thicker flakes showed a weaker gate response, likely due to electric field screening effects, with on/off ratios of ~10 (Supplementary Figure S2).

**TEMPERATURE DEPENDENT MOBILITY**

We measure these devices under "high" and "low" bias conditions, each probing a different transport regime. In the high bias regime, many states participate in transport, washing out any individual/few level transport effects. In this regime, we measure conductance ($G$) by applying a relatively large source-drain voltage ($V = 0.1$ V) as a function of backgate voltage ($V_{bg}$) at various temperatures and extract field effect mobility. Figure 2a displays a transconductance trace in a 12 nm thick Te flake device (device dimensions with width $W = 5.8$ μm and channel length $L = 11.5$ μm). By sweeping $V_{bg}$ from -60 V to 60 V, $G$ drops and pinches-off at ~ 5 V consistent with holes as the dominant charge carriers, exhibiting p-type behavior.[25,34] From the linear region of the transconductance trace, field effect mobility can be extracted using the equation:

$$\mu = \frac{L}{C_{ox} \cdot W} \frac{dG}{dV_{bg}}$$

where $C_{ox}$ is the gate capacitance.

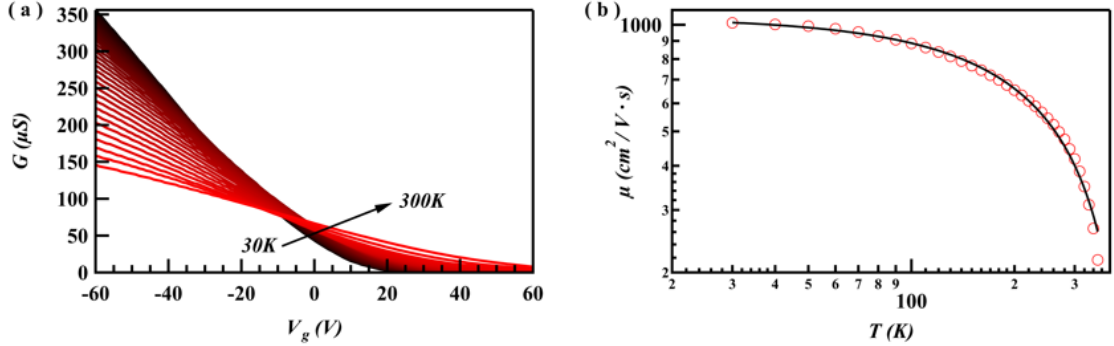

Figure 2. Temperature-dependent electrical transport properties of the device. (a) A 2D map of device conductance versus temperature and applied backgate voltage. (b) The extracted field-effect mobility as a function of temperature. At higher temperatures (rapid drop in mobility above T > 100 K), the mobility is limited by phonon scattering, as evidenced by its power law $T^{-\gamma}$. Conversely, the mobility saturates at a high value at low temperatures, indicating that scattering from charged impurities is not a dominant factor in this device.

The field effect mobility, calculated using the equation above, is depicted as a function of temperature in Figure 2b. In the phonon-limited transport regime, charge carrier mobility is expected to decrease with rising temperature according to the power law relation $\mu \sim T^{-\gamma}$. The value of the exponent, $\gamma$, distinguishes between different scattering mechanisms; acoustic phonon scattering typically results in $\gamma = 1$, whereas optical phonon scattering leads to $\gamma > 1$. Our analysis yielded a $\gamma$ value of approximately 1.26, indicating that optical phonons are the predominant scattering source limiting mobility in our tellurium device. This finding is consistent with similar observations in other two-dimensional material systems, such as $MoS_2$.[35,36] Additionally, the observed mobility saturation at low temperatures strongly suggests minimal influence from charge impurity scattering (up to 30K).[37] Mobility measured in another device is shown in Supplementary Figure S3, which is comparable to the previously reported values with similar Te thickness.[38]

Sn contacted devices exhibit ambipolar transport nature, due to narrow band gap in Te and a more left shifted threshold voltage for valence band transport. Conductance band transport appears at higher positive gate voltages. From the slopes of the transconductance characteristics, hole mobility was found to exceed electron mobility, possibly due to higher contact resistance for electrons arising from Schottky barriers at the conduction band compared to valance band (Supplementary Figure S4).

**QUANTUM TRANSPORT REGIME**

In the low-bias regime, the transport window narrows, restricting charge transport to a limited number of energy states. This selective probing enables the observation of fine quantum features that would otherwise be obscured at higher biases. At these low energies, comparable to or smaller than the sub-band spacings, quantum phenomena such as Coulomb blockade and interference effects become prominent.

Figure 3a presents the two-terminal conductance $G$ as a function of $V_{bg}$ measured at a base temperature in a dilution refrigerator (< 50 mK) in a device using Te thickness of 16 nm. Strong oscillations in $G$ are evident, indicating resonant features in the Te channel density of states. Two distinct features are observed. Figure 3b exhibits schematic of the mechanism in the device that gives rise to two features. To investigate these features further, we performed a bias spectroscopy, mapping $G$ as a function of source-drain bias $V$ and $V_{bg}$, as shown in Figure 3c. First, we observe diamond-shaped conductance domains visible near the pinch-off (low carrier density region), a characteristic of Coulomb blockade as shown in Figure 3d. These "Coulomb diamonds" arise from single-hole tunneling through a confined region weakly coupled to the leads (low barrier transparency). In this limit, hole transport is governed by Coulomb repulsion, enforcing sequential tunneling and resulting in sharply defined energy levels due to the increased lifetime of charge carriers (3b bottom schematic).

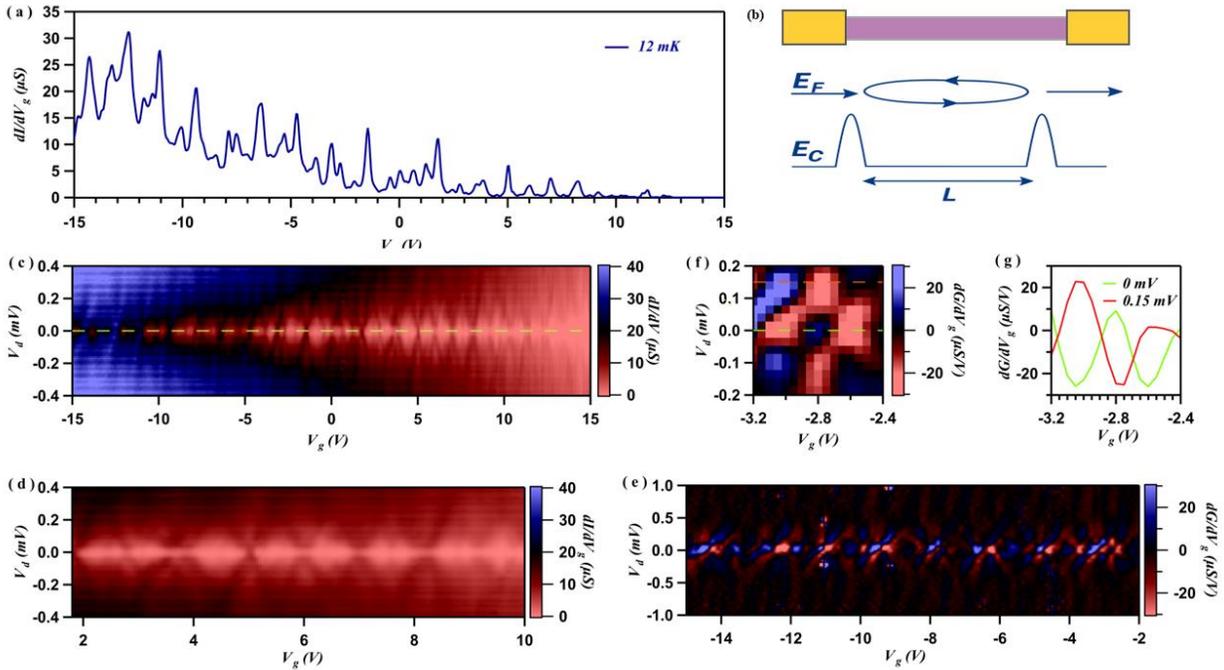

*Figure 3. Coulomb blockade and Fabry-Pérot interference in a hole gas. (a) Differential conductance measured in a thin Te device at a base temperature of 12 mK, displaying strong oscillations as the backgate voltage is swept. (b) A schematic illustrating the origin of the two distinct oscillatory phenomena observed. (c) A stability diagram plotting differential conductance as a function of source-drain bias and backgate voltage. The plot reveals two transport regimes: Coulomb blockade, characterized by Coulomb diamonds, dominates at low carrier density, while Fabry-Pérot interference emerges at higher carrier densities. (d) and (e) are magnified views of the regions in (c) dominated by Coulomb blockade and Fabry-Pérot interference, respectively. (f) A further zoomed-in view of the Fabry-Pérot regime, highlighting the checkerboard conductance pattern. (g) A line-cut from panel (f) demonstrates a phase shift of π in the conductance oscillations, a signature of alternating constructive and destructive interference.*

From the size of the Coulomb diamonds, we estimate a charging energy of approximately 400 μeV, corresponding to a self-capacitance of ~400 aF. From the capacitance we estimate the QD size of ~ 700 nm, fraction of the device dimension, suggesting the QD is formed within a fractional segment of the flake originating from potential fluctuations in the Te flake in the low carrier density region.

In the more negative $V_{bg}$, the second feature where $G$ pattern displays a checkerboard pattern (second derivative, Figure 3e) signifies an F-P interference regime,[33] where more transparent barriers are formed, and the discrete charge state (in case of Coulomb blockade) is no longer well-defined. In this open-dot regime, hole wavefunctions interfere constructively and destructively between two partially reflective barriers, forming a quantum cavity. This is more evident in the more negative $V_{bg}$ range (Figure 3b central schematic).

Systematically, the position of the $G$ peaks shifts linearly upon increasing $V$, as expected for Fabry-Perot interference.[39,40] The shift is also illustrated by Figure 4(f) (displaying a phase shift of $\pi$ due to constructive and destructive interference). This is clearly visible in a line cut shown in Figure 4g for measurements taken at $V = 0$ mV and 0.15 mV. From Figure 4(g), the bias needed to shift a maximum of differential conductance into a minimum is approximately 0.15 mV. The observed F-P resonances reflect the energy spacing $\Delta E$ between standing wave modes in the cavity. We observe different energy spacing in the whole $V_{bg}$ range measured from the observed checker-board pattern, ranging from $\Delta E = 0.1 - 0.4$ meV (Figure 4c and 4e). This suggests that there is possibly more than one cavity formed in our device. Using the energy spacing and cavity length relation,[33]

$$L_c = \frac{\hbar^2 \pi^2}{2m^* \Delta E}$$

we estimate a cavity length, $L_c$ of approximately 300 - 850 nm. F-P interference is observed when phase coherence length exceeds the length of the hole scattering centers (mean free path) forming the F-P cavity. Hence, the F-P cavity length provides a direct estimate of the mean free path in the system and lower bound on the phase coherence length. The phase coherence length scale extracted in our devices is consistent with prior reports on phase coherence length in similar Te systems from weak antilocalization measurments.[34]

Devices measured at low temperature in thicker Te flakes also exhibit F-P interference. However, we note that visibility of F-P in thicker flakes is significantly weaker compared to thinner flakes consistent with the picture that thinner flakes provide stronger confinement as compared to thicker flakes (Supplementary Figure S5).

**MAGNETIC FIELD EVOLUTION**

We now turn to the out-of-plane magnetic field ($B$) dependence in our device. Figure 4a displays a 2D color plot of $G$ evolution as a function of $V_{bg}$ and $B$. The data shown in Figure 4a is plotted after a polynomial background conductance subtraction to improve the $G$ peak evolution visibility. The unprocessed measured data is shown in Supplementary Figure S6.

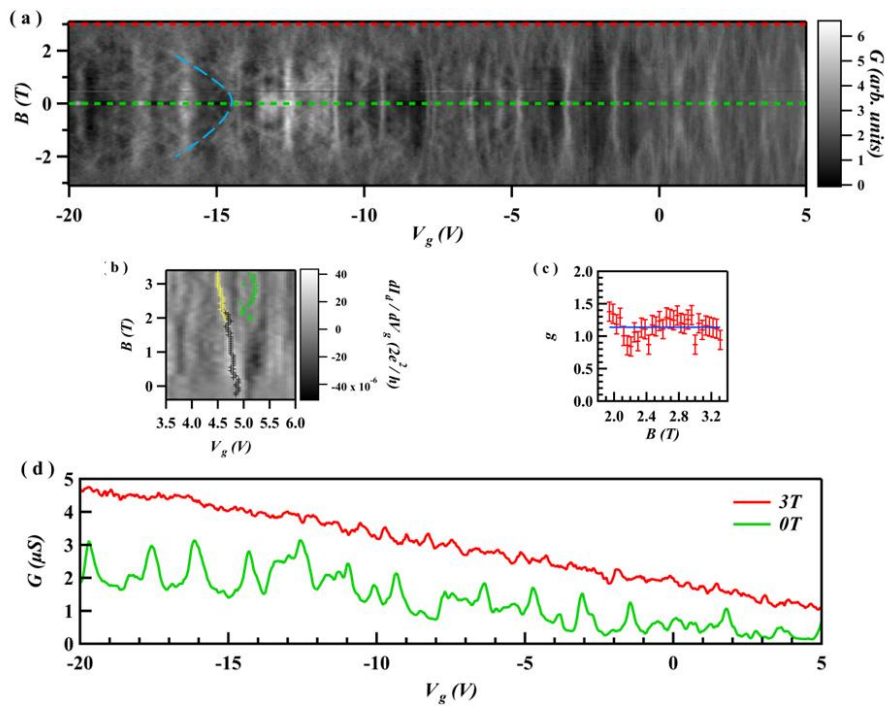

*Figure 4. Magneto-transport revealing distinct behaviors for Coulomb blockade and Fabry-Pérot interference. (a) Differential conductance plotted against backgate voltage and magnetic field. Two distinct evolutions are visible: linear dispersion of the Coulomb peaks at low carrier density and a non-linear dispersion of the Fabry-Pérot interference fringes at higher density, which bend significantly with the field (indicated by the blue dashed line). (b) An example of Zeeman splitting for a single Coulomb peak at a fixed backgate voltage of $V_{bg}$ =5 V. The splitting of the conductance peak is tracked using a software peak-finder. (c) The effective g-factor as a function of magnetic field, extracted from the Zeeman splitting shown in (b). (d) Line cuts of conductance traces taken at B=0 T and B=3 T. The prominent Fabry-Pérot oscillations observed at zero field completely disappear at 3 T. This suppression occurs when the cyclotron diameter (2Rc) becomes smaller than the Fabry-Pérot cavity length, effectively preventing the backscattering required for interference.*

The $G$ peak observed evolves in two ways. One set of $G$ peaks, particularly in the lower gate voltage range, exhibit a linear dispersion with $B$-field while some peaks exhibit bending of the oscillation and disappear above $B = 2$ T (Figure 4a). $G$ peaks that present a linear dispersion with $B$ field is a signature of charge localization and the linear shift corresponds to the energy level shift due to Zeeman splitting.[41] From the evolution of these energy levels with magnetic field (using $E = g\mu_B B$), we estimate a $g$-factor of 1.14. The experimentally determined g-factor in our device is smaller relative to the free-electron value ($g \approx 2$). A plausible explanation for this is the renormalization of the Te hole $g$-factor due to its hybridization with the superconducting leads (SnTe).[42] This suggests that the measured $g$-factor may describe the quasiparticle coherence peaks arising from the superconductor-semiconductor interface. A more detailed investigation is necessary to fully elucidate the underlying physical mechanism potentially using metal leads that are non-superconducting at low temperatures.

The F-P conductance oscillations, observed at large negative gate voltages, exhibit a non-linear dispersion with increasing magnetic field and are fully suppressed above $B \approx 2$ T. This behavior is attributed to the Lorentz force bending the hole trajectories within the cavity. The interference vanishes when the cyclotron diameter ($2R_c$) becomes smaller than $L_c$, which prevents back scattering and hence multiple reflections forming a standing wave.[41] At $V_{bg} = -15$ V, we estimate a Fermi velocity of $v_f \approx 6 \times 10^5$ m/s. This yields a cyclotron diameter of $2R_c \approx 336$ nm at the 2 T field, a value consistent with the cavity length of $L_c \approx 300\text{-}850$ nm estimated from the zero-field interference pattern (Figure 3d,e) with the upper limit of the $B$-field value determined by lower bound of the $L_c$. The red dashed line represents the magnetic field at which the classical cyclotron radius of the charge carriers exceeds the $L_c$.

**Conclusions:**

In summary, we fabricated and characterized field-effect devices based on thin tellurium flakes,

demonstrating their potential as a high-quality platform for quantum transport studies. We demonstrate a high carrier mobility in 17 nm thick flake reaching a value of 1000 cm²/V·s at 30K. At low temperatures, we observed a clear evolution from Coulomb blockade to Fabry-Pérot interference, with the visibility of these quantum oscillations being notably enhanced in thinner flakes. The application of a magnetic field allowed for the direct observation of Zeeman splitting.

The combination of high mobility and a gate-tunable quantum phenomena spikes interest for future explorations in thin tellurium flakes. The measurements also shed light on the device dimensions necessary to access the ballistic regime where individual sub-band transport could be explored. We expect a futher enhancement in carrier mobility and mean free path by encapsulating the flake in boron nitride. The results presented here pave the way for more complex device geometries and experiments aimed at harnessing the unique chiral and topological properties of tellurium. This includes investigating predicted Weyl physics, realizing topological superconductivity in hybrid devices, and developing novel low-power spintronic components.

**Methods:**

**SAMPLE PREPARATION**

Synthetic method of 2D Te nanosheets The synthesis of tellurene nanosheets was achieved through a sequential process utilizing metastable 1T'-MoTe$_2$ as dual-functional templates and Te sources in N-methylpyrrolidone (NMP) solvent. Initially, 1T'-MoTe$_2$ crystals (3 mg/mL) underwent prolonged bath ultrasonication (140 W, 10 h), followed by a 5-hour quiescent incubation period. The resultant grey translucent suspension was then homogenized through brief low-power sonication (60 W, 1 min) before undergoing a purification sequence involving primary centrifugation (3,000 rpm, 1 min), ethanol-assisted flocculation, and two dispersion-centrifugation cycles to yield stable tellurene colloids. For device fabrication, controlled deposition was executed by dropping the purified suspension onto SiO$_2$/Si substrates, followed by instantaneous nitrogen-flow drying to isolate individual nanosheets.

**DEVICE FABRICATION**

Device fabrication was carried out using a maskless photolithography writer to pattern the source and drain contacts in a two-step process. For the initial step, a bilayer photoresist was spin-coated onto the sample, and fine contacts designed to overlap the flake were defined. These contacts were then metallized by depositing Sn/Au or Pd/Au (20 nm/60 nm), followed by a standard lift-off procedure. Subsequently, the sample was coated again with the bilayer photoresist to pattern large-area contact pads connected to the fine leads. These pads were metallized with Cr/Au (5 nm/80 nm) to ensure robust connections for wire bonding.

**ACKNOWLEDGMENT**

This research is supported by MonArk NSF Quantum Foundry supported by National Science Foundation Q-AMASE-i program under NSF Award No. DMR-1906383 and AFRL under agreement number FA8750-24-1-1019. Q.H. acknowledges the substantial support from the Research Grants Council of Hong Kong (Projects No. 11314322).

TOC image

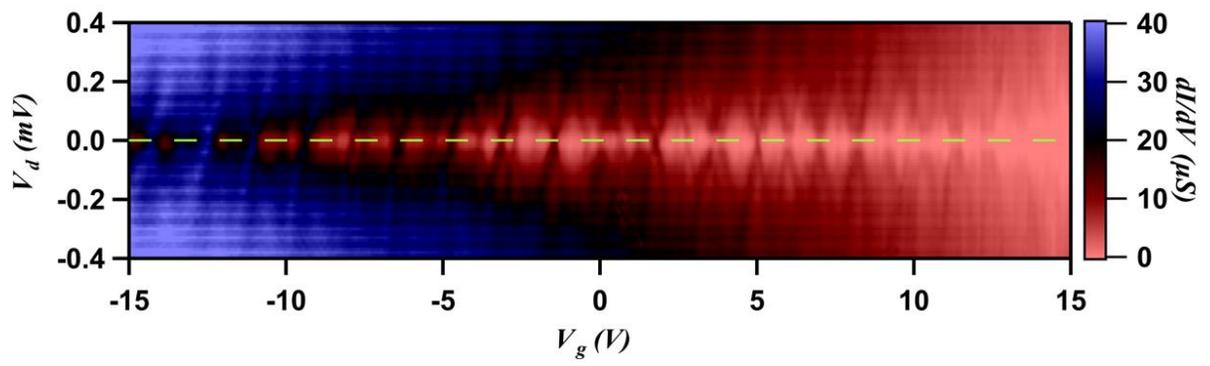

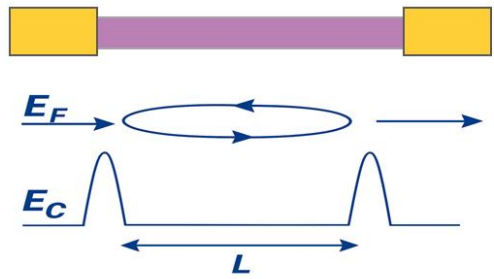

# Supplementary information for "Quantum Transport in Thin Two-Dimensional Tellurium Flakes"

*Mohammad Hafijur Rahaman[1], Nathan Sawyers[1], Mourad Benamara[2], Trudie Culverhouse[3,4], Repaka Maheswar[3], Qiyuan He[5], Hugh Churchill[1,6], Dharmraj Kotekar Patil[1,6]*

[1]Department of Physics, University of Arkansas, Fayetteville, AR 72701 USA

[2]Institute for Nano Science and Engineering, University of Arkansas, Fayetteville, AR 72701 USA

[3] Institute of Materials Research and Engineering, Agency for Science Technology and Research, (A*STAR) Singapore 138634, Republic of Singapore 138634

[4]Department of Chemistry, School of Natural Sciences, University of Manchester, Manchester M13 9PL, United Kingdom

[5]Department of Materials Science and Engineering, City University of Hong Kong, 83 Tat Chee Avenue, Kowloon, Hong Kong, China

[6]MonArk NSF Quantum Foundry, University of Arkansas, Fayetteville, AR 72701 USA

S1. AFM measurement in thick flakes
S2. Conductance on-off ratio in thick flake device
S3. Temperature dependent mobility in thick flake (20 nm)
S4. Ambipolar transport in Tin (Sn) contacted Te device
S5. Fabry-Perot interference in thick Te flake devices
S6. As measured magnetic field evolution of Figure 4a in the main text

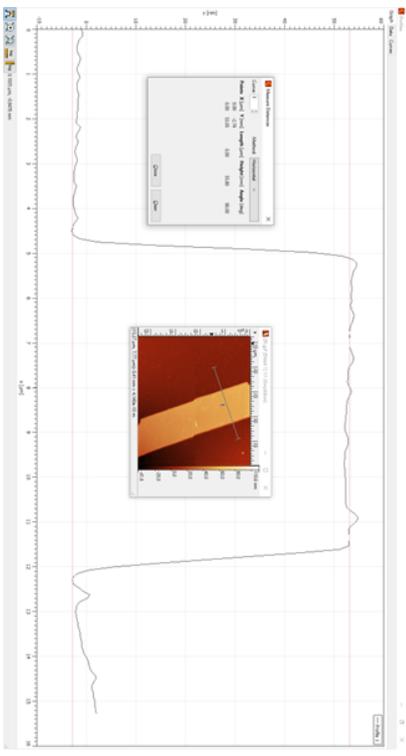
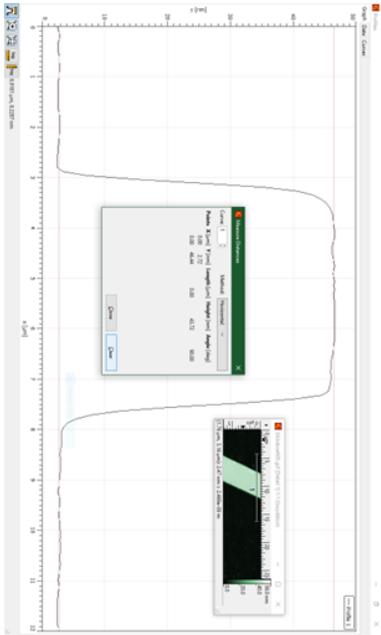
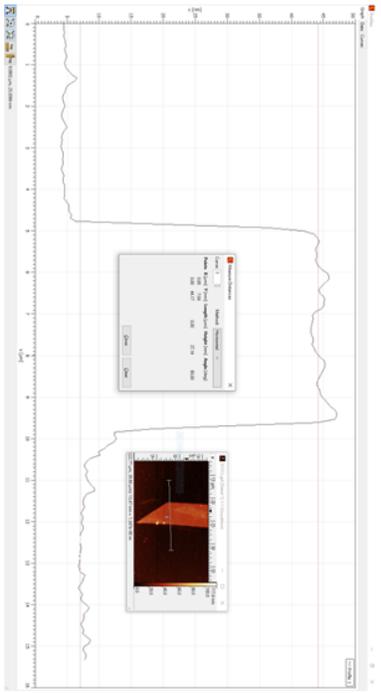

*S1:Atomic Force Microscope images of thick Te flakes.*

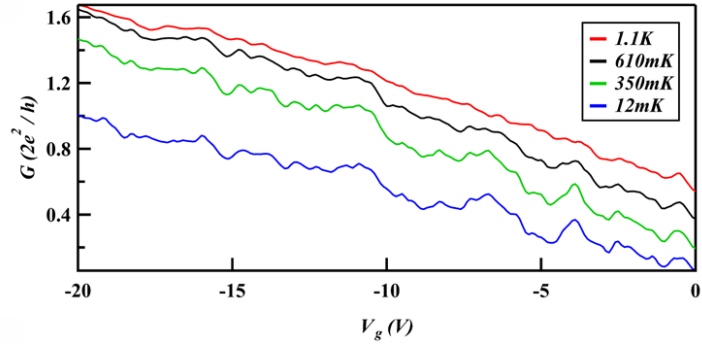

*S2: Line trace G vs $V_{bg}$ exhibiting weak gate dependence with on-off ratio of <10.*

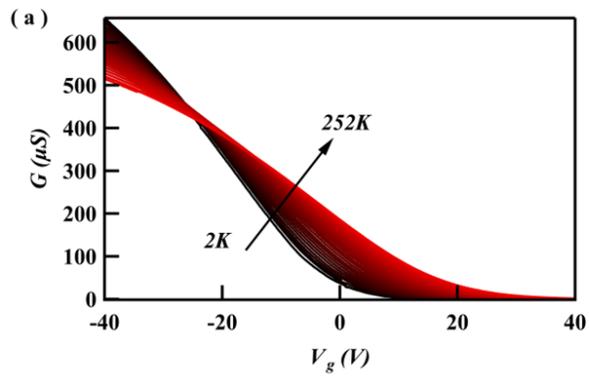 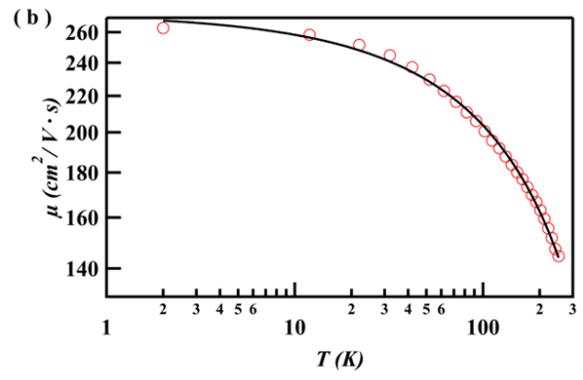

*S3: Conductance as a function of backgate and temperature in a Te device with thickness 20nm. Lower mobility is observed in this device compared to thinner flake device shown in the main text.*

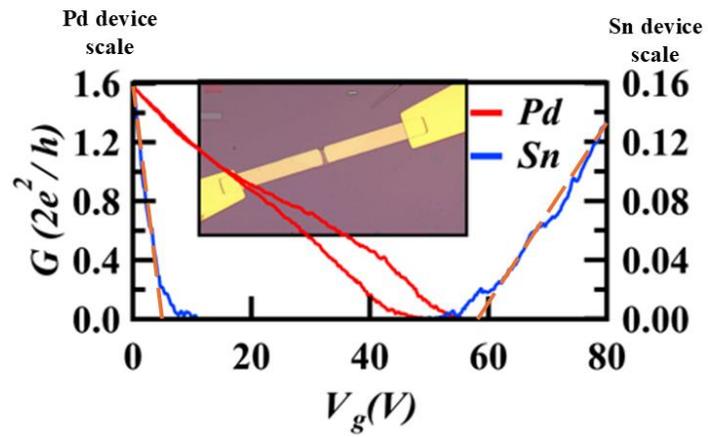

*S4: Comparison of Sn and Pd contacted Te flake device. Sn contacted Te device exhibits ambipolar behavior allowing access to conductance band. From the slope of transconductance trace (orange dashed line) hole mobility appears to be higher than the electron mobility.*

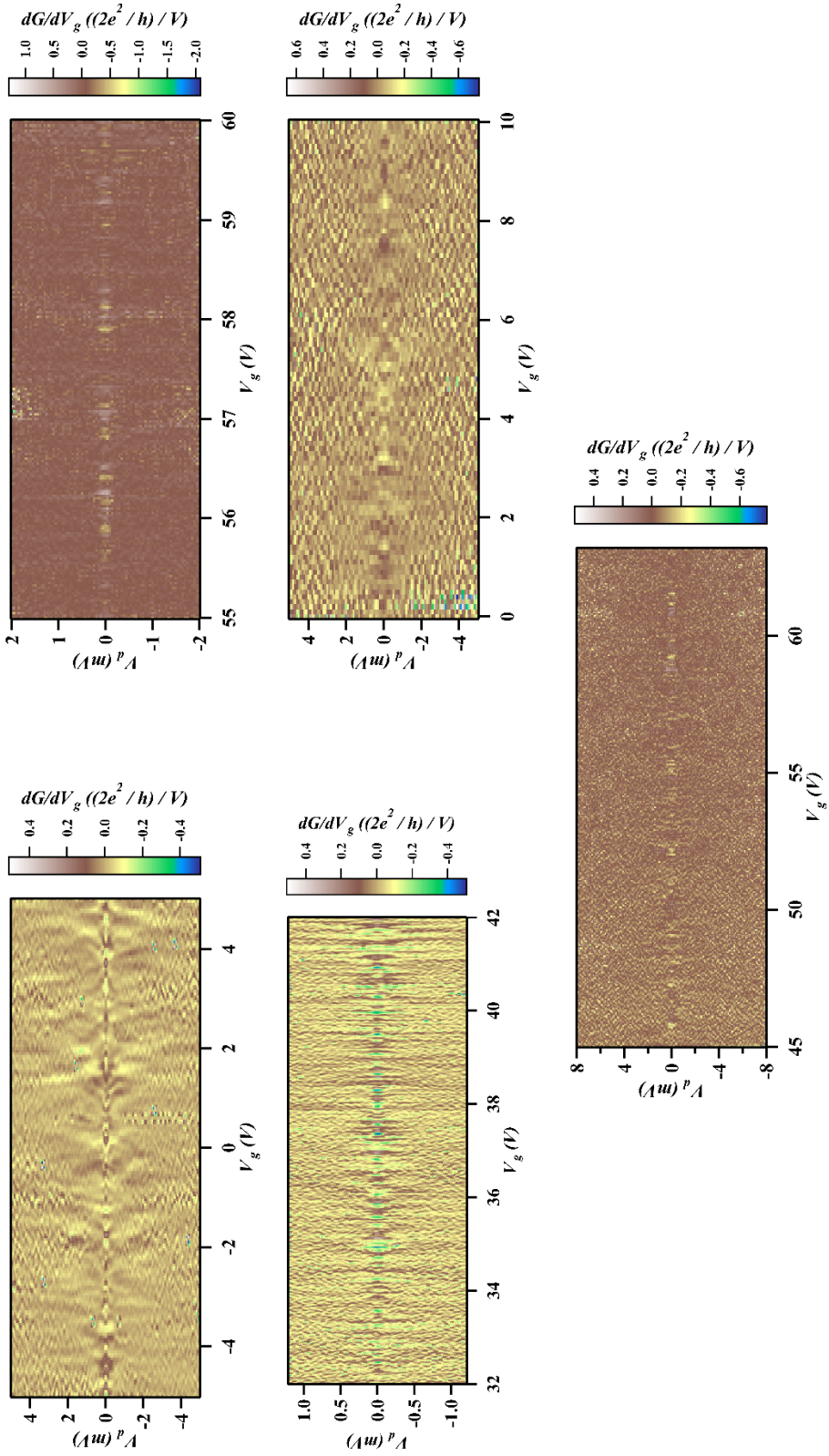

*S5: Low temperature characterization of thick Te flakes (thickness of 40 nm and above) exhibiting F-P interference. The observed F-P interference is significantly weak in thicker flakes as compared to thin flake shown in the main text.*

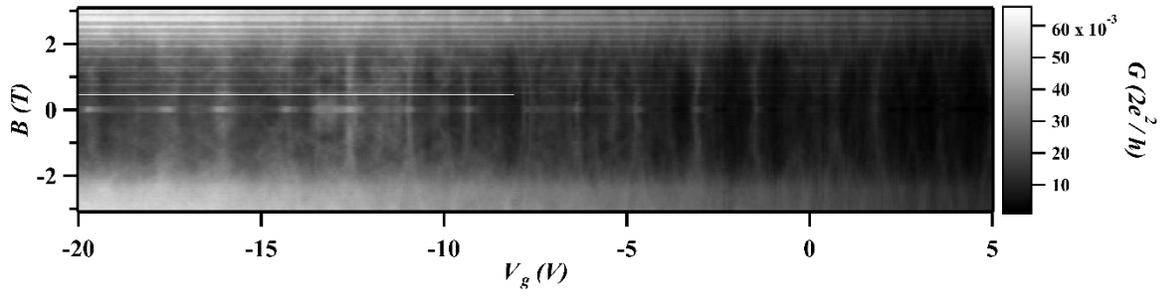

*S6: As measured conductance as a function of backgate and magnetic field for the data shown in figure 4a of the main text. The switching behavior and background is corrected for to display clear evolution of conductance peaks.*